\documentclass[aps,prl,superscriptaddress,twocolumn,showpacs,amsmath,amssymb,letter]{revtex4}

\usepackage{graphicx}
\usepackage{dcolumn}
\usepackage{bm}

\begin{document}

\title{Observation of Single Dirac Cone Topological Surface State in Compounds TlBiTe$_2$ and TlBiSe$_2$ from a New Topological Insulator Family}

\author{Yulin Chen}
\affiliation {Stanford Institute for Materials and Energy Sciences,
SLAC National Accelerator Laboratory, 2575 Sand Hill Road, Menlo
Park, California 94025}

\affiliation {Geballe Laboratory for Advanced Materials, Departments
of Physics and Applied Physics, Stanford University, Stanford,
California 94305}

\affiliation {Advanced Light Source, Lawrence Berkeley National
Laboratory Berkeley California, 94720, USA}

\author{Zhongkai Liu}
\affiliation {Stanford Institute for Materials and Energy Sciences,
SLAC National Accelerator Laboratory, 2575 Sand Hill Road, Menlo
Park, California 94025}

\affiliation {Geballe Laboratory for Advanced Materials, Departments
of Physics and Applied Physics, Stanford University, Stanford,
California 94305}

\author{James G. Analytis}
\affiliation {Stanford Institute for Materials and Energy Sciences,
SLAC National Accelerator Laboratory, 2575 Sand Hill Road, Menlo
Park, California 94025}

\affiliation {Geballe Laboratory for Advanced Materials, Departments
of Physics and Applied Physics, Stanford University, Stanford,
California 94305}

\author{Jiun-Haw Chu}
\affiliation {Stanford Institute for Materials and Energy Sciences,
SLAC National Accelerator Laboratory, 2575 Sand Hill Road, Menlo
Park, California 94025}

\affiliation {Geballe Laboratory for Advanced Materials, Departments
of Physics and Applied Physics, Stanford University, Stanford,
California 94305}

\author{Haijun Zhang}
\affiliation {Stanford Institute for Materials and Energy Sciences,
SLAC National Accelerator Laboratory, 2575 Sand Hill Road, Menlo
Park, California 94025}

\affiliation {Geballe Laboratory for Advanced Materials, Departments
of Physics and Applied Physics, Stanford University, Stanford,
California 94305}

\author{Sung-Kwan Mo}
\affiliation {Advanced Light Source, Lawrence Berkeley National
Laboratory Berkeley California, 94720, USA}

\author{Robert G. Moore}
\affiliation {Stanford Institute for Materials and Energy Sciences,
SLAC National Accelerator Laboratory, 2575 Sand Hill Road, Menlo
Park, California 94025}

\author{Donghui Lu}
\affiliation {Stanford Institute for Materials and Energy Sciences,
SLAC National Accelerator Laboratory, 2575 Sand Hill Road, Menlo
Park, California 94025}

\affiliation {Geballe Laboratory for Advanced Materials, Departments
of Physics and Applied Physics, Stanford University, Stanford,
California 94305}

\author{Ian Fisher}
\affiliation {Stanford Institute for Materials and Energy Sciences,
SLAC National Accelerator Laboratory, 2575 Sand Hill Road, Menlo
Park, California 94025}

\affiliation {Geballe Laboratory for Advanced Materials, Departments
of Physics and Applied Physics, Stanford University, Stanford,
California 94305}

\author{Shoucheng Zhang}
\affiliation {Stanford Institute for Materials and Energy Sciences,
SLAC National Accelerator Laboratory, 2575 Sand Hill Road, Menlo
Park, California 94025}

\affiliation {Geballe Laboratory for Advanced Materials, Departments
of Physics and Applied Physics, Stanford University, Stanford,
California 94305}

\author{Zahid Hussain}
\affiliation {Advanced Light Source, Lawrence Berkeley National
Laboratory Berkeley California, 94720, USA}

\author{Z.-X. Shen}
\affiliation {Stanford Institute for Materials and Energy Sciences,
SLAC National Accelerator Laboratory, 2575 Sand Hill Road, Menlo
Park, California 94025}

\affiliation {Geballe Laboratory for Advanced Materials, Departments
of Physics and Applied Physics, Stanford University, Stanford,
California 94305}

\date{\today}

\begin{abstract}
Angle resolved photoemission spectroscopy (ARPES) studies were
performed on two compounds (TlBiTe$_2$ and TlBiSe$_2$) from a
recently proposed three dimensional topological insulator family in
Thallium-based III-V-VI$_2$ ternary chalcogenides. For both
materials, we show that the electronic band structures are in broad
agreement with the $ab$ $initio$ calculations; by surveying over the
entire surface Brillouin zone (BZ), we demonstrate that there is a
single Dirac cone reside at the center of BZ, indicating its
topological non-triviality. For TlBiSe$_2$, the observed Dirac point
resides at the top of the bulk valance band, making it a large gap
($\geq$200$meV$) topological insulator; while for TlBiTe$_2$, we
found there exist a negative indirect gap between the bulk
conduction band at $M$ point and the bulk valance band near
$\Gamma$, making it a semi-metal at proper doping. Interestingly,
the unique band structures of TlBiTe$_2$ we observed further suggest
TlBiTe$_2$ may be a candidate for topological superconductors.

\end{abstract}


\maketitle

Topological insulators represent a new state of quantum matter with
a bulk gap and odd number of relativistic Dirac fermions on the
surface \cite{1}. Since the discovery of two dimensional (2D)
topological insulator in HgTe quantum well \cite{2,3} and subsequent
in three dimensional (3D) materials (especially the single Dirac
cone family Bi$_2$Te$_3$, Bi$_2$Se$_3$ and
Sb$_2$Te$_3$)\cite{11,12,13}, topological insulators has grown as
one of the most intensively studied fields in condensed matter
physics \cite{1,2,3,11,12,13,4,5,8,9,10}. The massless Dirac
fermions and the magnetism further link the topological insulators
to relativity and high energy physics \cite{13.1}. The fast
development of the topological insulators also inspires the study of
other topological states such as topological superconductors
\cite{14,15,16,17,18,26,27}, which has a pairing gap in the bulk and
topologically protected surface state consisting of Majorana
fermions \cite{14}. Unlike Dirac fermions in topological insulators
that can have the form of particles or holes, Majorana fermions are
their own antiparticles \cite{19}. The simplest 3D topological
superconductor consists of a single Majorana cone on the surface,
containing half the degree of freedom of the Dirac surface state of
a single cone 3D topological insulator. This fractionalization of
the degree of freedom introduces quantum non-locality and is
essential to the topological quantum computing based on Majorana
fermions \cite{20}.

In this work, we use ARPES to study the electronic structure of
TlBiTe$_2$ and TlBiSe$_2$ from a recently proposed topological
insulator family: Thallium-based III-V-VI$_2$ ternary chalcogenides
\cite{21, 22}. Remarkably, both the surface and bulk electronic
structures are in broad agreement with the $ab$ $initio$
calculations; and a single Dirac cone centered at the $\Gamma$ point
of surface Brillouin zone is found in both materials. Furthermore,
for the p-type TlBiTe$_2$, the experimental band structure shows six
leaf-like bulk valence band pockets around the Dirac cone. Given
that these leaf-like bulk pockets are the only structure other than
the surface Dirac cone on the Fermi-surface (FS), they may provide a
possible origin of the reported bulk superconductivity \cite{23},
which can further induce superconductivity on the surface state by
proximity effect, making TlBiTe$_2$ a candidate for 3D topological
superconductors. Another compound of the family, TlBiSe$_2$ has a
simpler bulk structure around the single Dirac cone at the $\Gamma$
point, with the Dirac point resides on top of the the bulk energy
gap ($\sim$200meV), making it a large gap topological insulator
similar to Bi$_2$Se$_3$ \cite{11}, but with better mechanical
properties than the Bi$_2$Se$_3$/Bi$_2$Te$_3$ family, as the bonding
between layers are much stronger \cite{21} than the van de Waals'
force that bonds quintuple layer units of Bi$_2$Te$_3$ or
Bi$_2$Se$_3$ \cite{11}.

\begin{figure}
\includegraphics[width=0.48\textwidth]{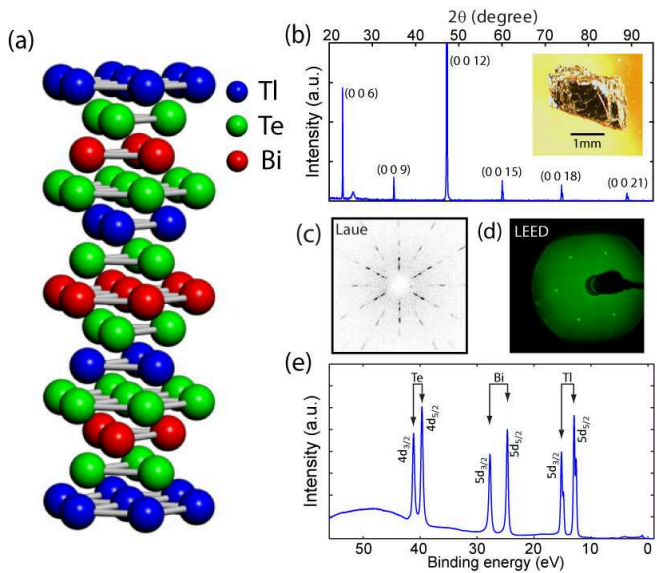}
\caption{\label{fig:epsart} (color) (a) Crystal structure of
TlBiTe$_2$ with repeating layers -Tl-Te-Bi-Te-. (b) XRD study on the
as grown crystal shows clean characteristic peaks. The photo of a
sample in inset shows mirror-like cleavage plane parallel to the
(111) plane. (c) Sharp Laue pattern confirms the high quality of the
crystals used for ARPES measurements. (d) LEED pattern on a cleaved
surface after ARPES measurement demonstrates clear surface
diffraction spots without surface reconstruction. (e) Core level PES
demonstrates the characteristic peaks from d-shell electrons of all
three compositional elements.}
\end{figure}

The crystal structure of Thallium based III-V-VI$_2$ ternary
chalcogenides is rhombohedra with the space group R-3m, which can be
viewed as a distorted NaCl structure with four atoms in the
primitive unit cell. A conventional unit cell of TlBiTe$_2$ is shown
in Fig. 1(a) as an example: the three different types of atoms stack
in layers with repeating sequence ...-Te-Bi-Te-Tl-...\cite{24}. The
existence of a flat cleavage plane \cite{25} parallel to the (111)
surface in this family of compounds [inset of Fig. 1(b)] makes them
suitable for ARPES study. The high quality of the crystal was
demonstrated by the XRD [Fig. 1(b)] and Laue [Fig 1(c)]
characterizations; and the LEED pattern on the sample surface after
ARPES measurement (Fig. 1d) confirmed that the surface structure of
the samples is free from reconstruction. The characteristic peaks of
all three elements can be identified in core level photoemission
measurements [Fig.1(e)].

\begin{figure*}
\includegraphics[width=\textwidth]{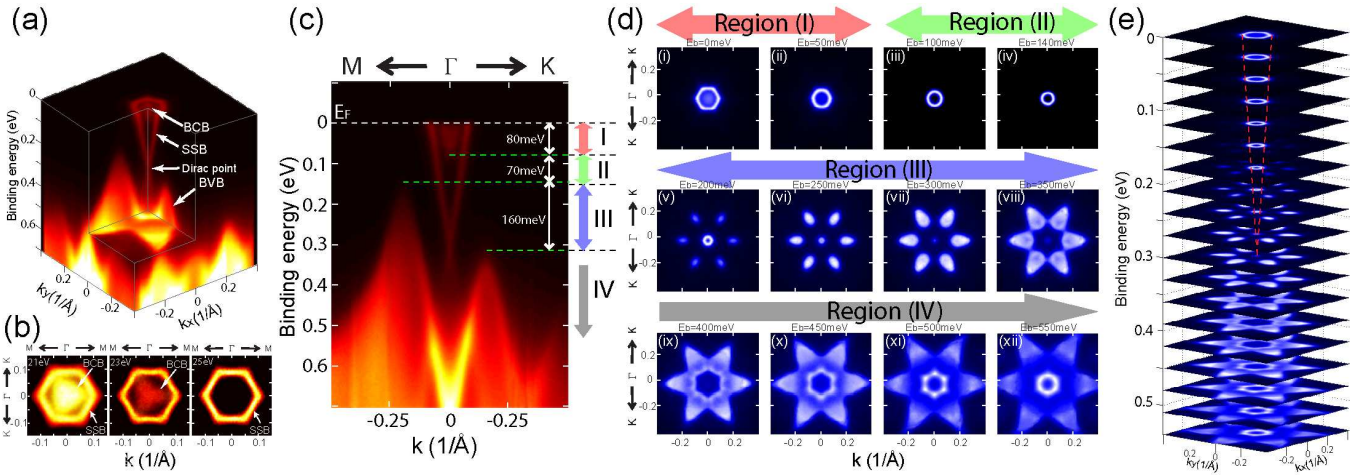}
\caption{\label{fig:epsart} (Color) (a) 3D representation of the
dispersion, with the bulk conduction band (BCB), bulk valance band
(BVB) surface state band (SSB) and the Dirac point indicated. (b)
Photon energy dependent FS maps (symmetrized according to the
crystal symmetry) shows different BCB pockets for 21, 23 and 25eV
photons. (c) Band dispersions along the M-$\Gamma$-K direction. Four
regions defined by characteristic energy positions of the band
structure are labeled. (d) Constant energy plot of the band
structure in different regions defined in (c), showing the BCB FS
inside SSB (region I), SSB only (region II), BVB outside SSB (region
III) and all BVB (region IV). (e) Stacking constant energy plots
illustrates the evolution of the band structure in different
regions. Red dashed line traces the dispersion of the SSB from the
Dirac point.}
\end{figure*}

Figure 2 displays TlBiTl$_2$ band structure around the center of the
BZ. The 3D band structure [Fig.2(a)] shows a clear Dirac cone
centered at the $\Gamma$-point with broad features from bulk
conduction (BCB) and bulk valence band (BVB) as theoretically
predicted \cite{21,22}. To confirm the surface nature of the Dirac
cone, excitation photon energy dependent ARPES study (Fig.2b) was
performed. The non-varying shape of the outer hexagonal surface
state band (SSB) FS with different excitation photon energies
indicates its 2D nature; while the shape and the existence of the
BCB FS pocket inside changes dramatically due to its 3D nature with
strong $k_z$ dispersion as expected.

Detailed band dispersions along two high symmetry directions
($\Gamma-M$ and $\Gamma-K$) are illustrated in Fig. 2(c), in which
an asymmetry of the BVB along the two directions can be seen. Based
on the characteristic energy positions of the bulk band, we can
divide the band structure into four regions [Fig. 2(c)] for
discussion of different FS geometries [Fig. 2(d)]. From region I to
III, the bulk contribution of the FS evolves from an n-type pocket
inside (region I) the n-type SSB FS to six p-type leaf-like pockets
outside (Region III), with the bulk pockets disappear in region II;
while in region IV, both the FS at the center and the surrounding
leaf-like pockets are p-type. Fig. 2(e) shows a summary of the band
evolution through all four regions.

\begin{figure*}
\includegraphics[width=\textwidth]{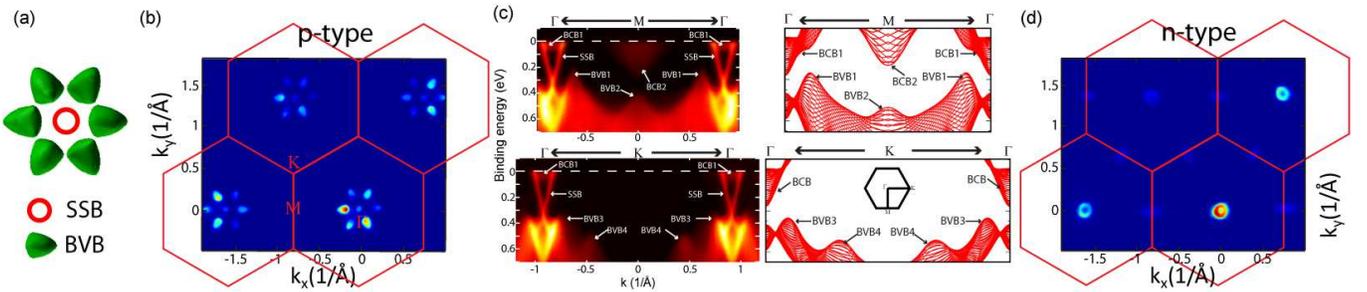}
\caption{\label{fig:epsart} (Color) (a)Illustration of the typical
FS in region III as defined in Fig. 2. The SSB pocket at $\Gamma$ is
surrounded by six leaf-like BVB FS pockets. (b) Broad k-space scan
covers four BZs shows no additional features besides the FS pockets
illustrated in (a), with the $\Gamma$ , M and K points marked. (c)
Comparison between the measured band structure (left sub panels) and
$ab$ $initio$ calculations (right sub-panels). Results of
$\Gamma-M-\Gamma$ ($\Gamma-K-\Gamma$) direction are shown on the top
(bottom) row. Prominent BCB, BVB and SSB features are marked in both
measured and calculated band structures. (d) Broad FS map of the
n-type sample shows additional BCB pockets at M point in addition to
the ring-like SSB pocket around $\Gamma$, as a result from the BCB2
band in panel (c) (top row).}
\end{figure*}

Besides having the single Dirac cone on the surface, TlBiTe$_2$ was
also reported to superconduct when p-doped \cite{23}, with $E_F$ of
the corresponding density ($\sim6\times10^{20}/cm^3$) resides in
region III (about 150meV below the bottom of BVB). From our
measurements, the FS geometry in this region is characterized by a
ring like SSB FS and six surrounding p-type bulk pockets [Fig.
3(a)], as clearly shown in Fig. 2(d) - where a broad scan in
$k$-space that covers four BZs [Fig. 3(b)] confirms that the FS
structure in Fig. 3(a) is the only feature within each BZ. This
leads to the natural conclusion that the bulk superconductivity of
p-type TlBiTe$_2$ originates from the six leaf-like bulk pockets;
and in the superconducting state, the surface state [the center FS
pocket in Fig. 3(a,b)] can become superconducting due to the
proximity effect induced by the bulk states. For such a
superconductor, it has been proposed \cite{14} that each vortex line
has two Majorana zero modes related by the time reversal symmetry,
thus making it a candidate for topological superconductors and
suitable for the topological quantum computation \cite{20}. However,
the presence of superconductivity in p-type TlBiTe$_2$ requires
further confirmation \cite{23.1, 23.2}.

The band structures of TlBiTe$_2$ in larger energy and momentum
scale are shown in Fig. 3(c), where the measured dispersions (left
sub-panels) along both $\Gamma-M-\Gamma$ and $\Gamma-K-\Gamma$
directions are compared with the corresponding $ab$ $initio$
calculation of the bulk band (right sub-panels). In general, the
experimental dispersions along both directions agree well with the
calculation, which reproduces each bulk feature of the measurement
[Fig. 3(c)], albeit the relative energy position is slightly
different. Again, the non-existence of the linear dispersion of the
Dirac cone in the $ab$ $initio$ bulk calculation confirms its
surface nature.

Interestingly, from the measurements [Fig. 3(c), top left panel], we
find there is a small energy overlap ($\sim$20meV) between the
bottom of the electron pocket at M (BCB2) and the top of the valence
band around $\Gamma$ (BVB1), indicating that TlBiTe$_2$ is a
semi-metal if $E_F$ resides in this region. Also, unlike the p-type
sample, the FS of an n-type TlBiTe$_2$ (electron density
$\sim10^{19}/cm^3$) shows an electron pocket at the M point [Fig.
3(d)] due to the bulk conduction band (BCB2) at the zone boundary
[Fig. 3(c), top row].

\begin{figure*}
\includegraphics[width=\textwidth]{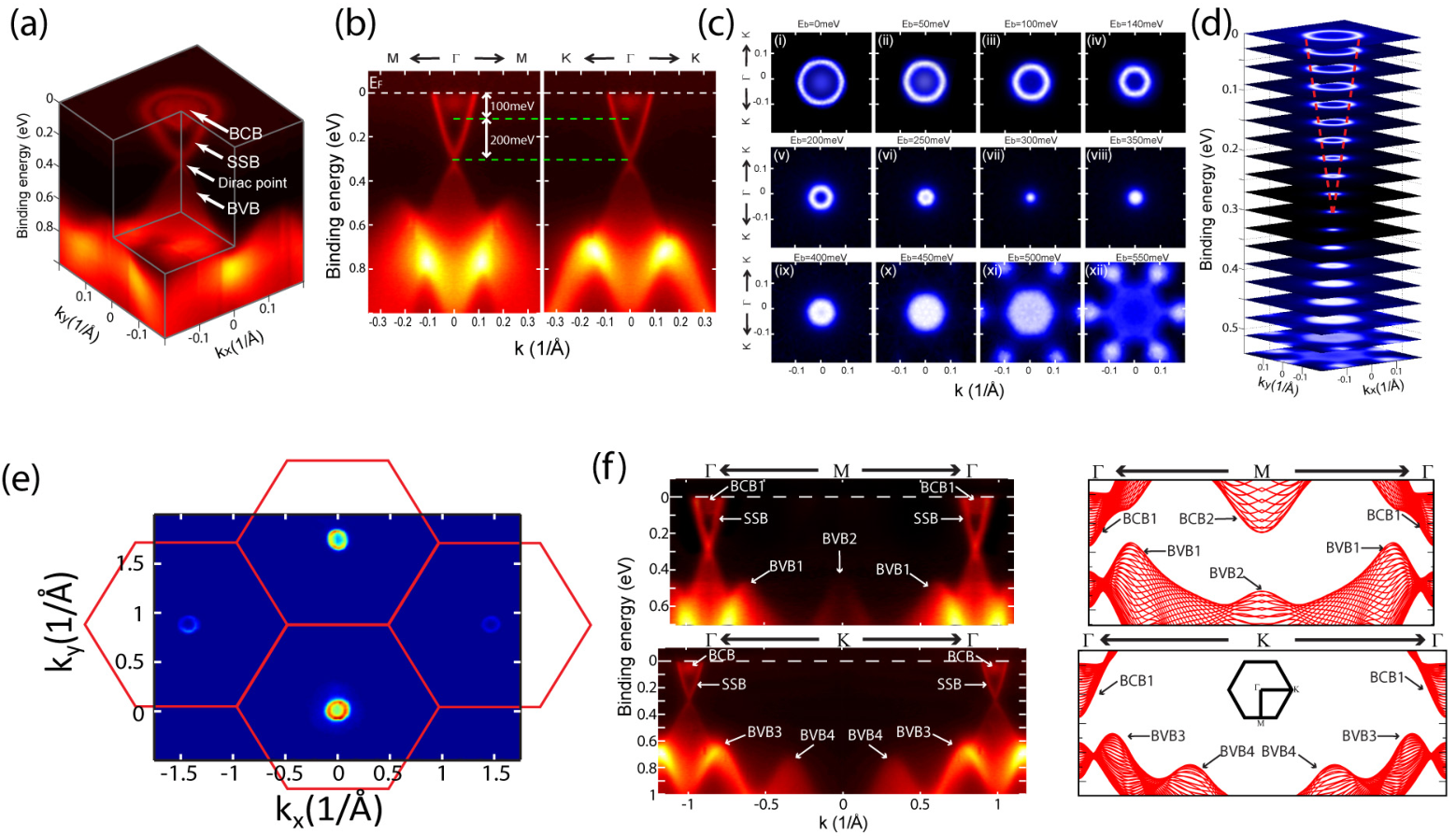}
\caption{\label{fig:epsart} (Color) (a) 3D illustration of the band
structure around $\Gamma$, with the BCB, BVB, SSB and the Dirac
point indicated. (b) Detailed band structure along M-$\Gamma$-K
direction, with less anisotropy compared to Fig. 2(c). The bottom of
the BCB at $\Gamma$ is about 100meV below $E_F$ and the Dirac point,
residing on top of the BVB, is about 200meV below the BCB bottom.
(c) Constant energy contours of the band structure at different
binding energies. (d) Stacking constant energy plots shows the
evolution of the band structure. Red dashed line traces the
dispersion of the SSB from the Dirac point. (e) Broad FS map of
n-type TlBiSe$_2$ shows a FS sheet without BCB pocket at M point.
(f) Comparison between the measured band structure (left sub panels)
and calculations (right sub-panels). Results of $\Gamma-M-\Gamma$
($\Gamma-K-\Gamma$) direction are shown on the top (bottom) row.
Prominent BCB, BVB and SSB features are marked in both the measured
and calculated band structures. }
\end{figure*}

The band structure of TlBiSe$_2$, another compound from the Tl-based
ternary family, is summarized in Fig. 4. Similar to TlBiTe$_2$,
there exists a single surface Dirac cone at the $\Gamma$ point [Fig.
4(a-f)], confirming that its topological non-triviality. The main
difference between TlBiSe$_2$ and TlBiTe$_2$ is that the Dirac point
of TlBiSe$_2$ resides at the top of the BVB [Fig. 4(a,b)], and the
system has a $\sim$200meV direct bulk gap at $\Gamma$. The bulk band
structure is also simpler around $\Gamma$ and less anisotropic along
$\Gamma$-M and $\Gamma$-K directions [Fig. 4(a,b)]. This simplicity
is also shown in the constant energy contour plots [Fig. 4(c)] and
its evolution [Fig. 4(d)]. Comparing Fig. 4(c) and Fig. 2(b), we
notice that the SSB FS of TlBiSe$_2$ is a convex hexagon, contrast
to that of TlBiTe$_2$ which shows slightly concave geometry. This
difference resembles the difference between the SSB FSs of
Bi$_2$Te$_3$ and Bi$_2$Se$_3$, and can be reflected by different
observations in experiments such as scanning tunneling
microscopy/spectroscopy STM/STS \cite{28,29}.

Besides the simpler band geometry around $\Gamma$, the broad range
FS map [Fig. 4(e)] of n-type TlBiSe$_2$ is also simpler than that of
the TlBiTe$_2$ [Fig. 3(d)], without the electron pocket at the M
point. This simplicity can also be seen in the band dispersions in
Fig. 4(f), where although the experimental (left sub-panels) and
calculated (right sub-panels) bulk band structure again show
agreement in general, the BCB2 feature at M in the calculation (top
right panel) was not seen in the measurements (top left panel),
causing the missing of an electron pocket at M in Fig. 4e.

\bigskip
\textbf{Acknowledgements} We thank X. L. Qi, B.H. Yan and C.X. Liu
for insightful discussions. This work is supported by the Department
of Energy, Office of Basic Energy Science under contract
DE-AC02-76SF00515.

\end{document}